\documentclass[a4paper,preprint,preprintnumbers,amsmath,amssymb,superscriptaddress]{revtex4}

 \usepackage[T1]{fontenc}
 \usepackage{array}
 \usepackage{amsmath}
\usepackage{tabularx}
 \usepackage{graphicx}
\usepackage{multirow}
 \usepackage{SIunits}
 \usepackage{ulem}
 \usepackage{braket}
 \usepackage{color}
 \usepackage[draft]{fixme}
\usepackage{amssymb}
\usepackage{dcolumn}
\usepackage{bm}
\usepackage{footmisc}
\usepackage{hyperref}

\newcolumntype{P}[1]{>{\centering\arraybackslash}p{#1}}

\begin{document}

\begin{abstract}

We report on the optical properties of single InAs/GaAs quantum dots emitting near the telecommunication O-band, probed via Coulomb blockade and non-resonant photoluminescence spectroscopy, in the presence of external electric and magnetic fields. We extract the physical properties of the electron and hole wavefunctions, including the confinement energies, interaction energies, wavefunction lengths, and $g$-factors. For excitons, we measure the permanent dipole moment, polarizability, diamagnetic coefficient, and Zeeman splitting. The carriers are determined to be in the strong confinement regime. Large range electric field tunability, up to 7 meV, is demonstrated for excitons. We observe a large reduction, up to one order of magnitude, in the diamagnetic coefficient when rotating the magnetic field from Faraday to Voigt geometry due to the unique dot morphology. The complete spectroscopic characterization of the fundamental properties of long-wavelength dot-in-a-well structures provides insight for the applicability of quantum technologies based on quantum dots emitting at telecom wavelengths.

\end{abstract}

\title{Magneto-optical spectroscopy of single charge-tunable InAs/GaAs quantum dots emitting at telecom wavelengths}

\author{Luca Sapienza}\thanks{Equal contribution}\email{l.sapienza@soton.ac.uk}
\affiliation{Department of Physics and Astronomy, University of Southampton, Southampton SO17 1BJ, United Kingdom}
\author{Rima Al-Khuzheyri}\thanks{Equal contribution}
\author{Adetunmise Dada}
\affiliation{Institute of Photonics and Quantum Sciences, SUPA, Heriot-Watt University, Edinburgh, United Kingdom}
\author{Andrew Griffiths}
\author{Edmund Clarke}
\affiliation{EPSRC National Centre for III-V Technologies, University of Sheffield, United Kingdom}
\author{Brian D. Gerardot}
\affiliation{Institute of Photonics and Quantum Sciences, SUPA, Heriot-Watt University, Edinburgh, United Kingdom}
\date{\today}

\maketitle

\section{Introduction}
Single quantum dots grown by molecular beam epitaxy are one of the most promising sources of non-classical light due to their stable and sharp emission lines and easy integration on a chip via the mature III-V semiconductor fabrication technology. 
In particular, InAs quantum dots emitting at wavelengths around 950 nm have proved to be pure sources of single indistinguishable photons \cite{Reiz_ind_photons, Ding} and entangled photons \cite{entang, entang2} and a powerful platform for spin initialization and manipulation, spin-photon and remote spin entanglement \cite{QDspin_review}. To encode information in single photons and transmit it over long distances, sources of quantum light emitting at the so-called telecommunication wavelengths are most desirable. Advances in the development of superconducting detectors operating at cryogenic temperatures \cite{sup_det} allow detection efficiencies exceeding 90\% \cite{sup_det2}, making single-photon experiments and technologies eminently feasible. The growing interest in the field of long wavelength quantum dots is demonstrated by recent achievements such as the demonstration of bright sources of indistinguishable photons \cite{Edo}, interference of photons emitted by dissimilar sources \cite{Shields_inter}, entangled photon pair generation \cite{Shields_entang} and exciton fine-structure splitting manipulation \cite{PRB} in the telecom wavelength band. 

However, the growth and fundamental characterisation of quantum dots emitting at telecom wavelengths is less mature compared to emitters at wavelengths $<$ 1 $\mu$m. The longer emission wavelength can be achieved by growing quantum dots in a quantum well (the so-called dot-in-a-well or DWELL structures \cite{DWELL}), a technique that partially relaxes the strain accumulated during the Stranski-Krastanow growth, resulting in larger quantum dot dimensions. The InGaAs quantum well provides local strain relief and also preserves the quantum dot composition and height during growth by reducing the out-diffusion of In during capping of the dot layer \cite{DWELL1, DWELL2}. A larger physical confining potential implies a reduced energy separation between the quantum dot confined states and radiative electron-hole recombinations for InAs/GaAs quantum dots can reach wavelengths around 1.3 $\mu$m both at room temperature, for example for quantum dot lasers \cite{laser}, and at low temperature for single-quantum dot applications \cite{QD}. Besides the technological interest associated with lower transmission losses, telecom-wavelength quantum dots present interesting fundamental properties due to very different confinement of electron and hole wavefunctions and potentially larger oscillator strengths compared to shorter-wavelength quantum dots. In order to translate the more developed technology of 950 nm-band quantum dots towards longer wavelengths, the fundamental properties of the emitters need to be further understood. To this end, the application of external electric and magnetic fields as well as Coulomb blockade is a means to characterise the electron and hole wavefunctions and the Coulomb interactions between carriers. Since quantum dots emitting around 1300 nm are physically larger and have a higher In-composition than shorter wavelength quantum dots, the different composition and morphology can result in a different wavefunction extensions and electron-hole overlap,  impacting their fundamental response to applied fields. In this direction, analysis of the emission properties of quantum dots emitting at wavelengths > 1.2 $\mu$m in the presence of an external magnetic field \cite{1550_B, Kleemans, Cade} and of 1300 nm quantum dots in the presence of external strain \cite{PRB} have been reported. However, the full characterisation of the fundamental properties of quantum dots allowing direct comparison of emitters at 950 nm and at telecom wavelengths is still incomplete.

Here, we report on magneto-optical studies of the emission properties of single telecom-wavelength quantum dots grown within a charge-tunable structure. We extract the physical properties of the electron and hole wavefunctions, including the confinement energies, interaction energies, wavefunction lengths, and $g$-factors. For excitons, we measure the permanent dipole moment, polarizability, diamagnetic coefficient, and Zeeman splitting.

\section{Experimental Details}

The sample investigated was grown by molecular beam epitaxy. The DWELL layer was grown at 500$^{\circ}$C by initial deposition of 1 nm In$_{0.18}$Ga$_{0.82}$As, followed by a deposition of nominally 1.8 monolayers (ML) of InAs to form the quantum dots, at a growth rate of 0.016 MLs$^{-1}$. Sample rotation was stopped during growth of the quantum dot layer to provide a variation in InAs coverage across the wafer, resulting in a variation in quantum dot density across the wafer. The quantum dot layer was subsequently capped by 6 nm In$_{0.18}$Ga$_{0.82}$As and a further 4 nm GaAs at 500$^{\circ}$C, before the substrate temperature was raised to 580$^{\circ}$C for growth of the remaining structure. A cross-section transmission electron microscopy (TEM) image of a DWELL layer grown under similar conditions is shown in Ref.\cite{PRB}. Analysis of TEM images indicates the dots have a base width of 20-30 nm and a relatively large capped height of around 8-10 nm, preserved due to reduced out-diffusion of In during capping by the InGaAs layer.
A sketch of the energy diagram of the field effect structure with a single layer of DWELL quantum dots is shown in Fig. 1a. By applying a voltage between the semi-transparent NiCr Schottky gate  on the sample surface and the doped ($n^{+}$) GaAs layer (Ohmic contact), discrete charging of the dots with single electrons can be achieved \cite{charge_tun}. The sample is placed in a cryostat at $\sim$ 4 K and, using a microscope in confocal geometry, a fiber-coupled non-resonant (830 nm-wavelength) laser is used to excite the emitters. A zirconia super solid-immersion lens is positioned on the surface of the sample to reduce the excitation spot (and therefore be able to excite single emitters in the relatively high density sample) and increase the collection efficiency of the photoluminescence signal \cite{SIL}. The emission from single quantum dots is then coupled to a single-mode fiber and sent into a grating spectrometer equipped with an InGaAs array detector for spectral characterisation. An external magnetic field can then be applied to the sample either parallel or orthogonal to the growth axis.

\begin{figure}[tb]
   \centering
  \includegraphics[height=18cm, width=0.75\linewidth]{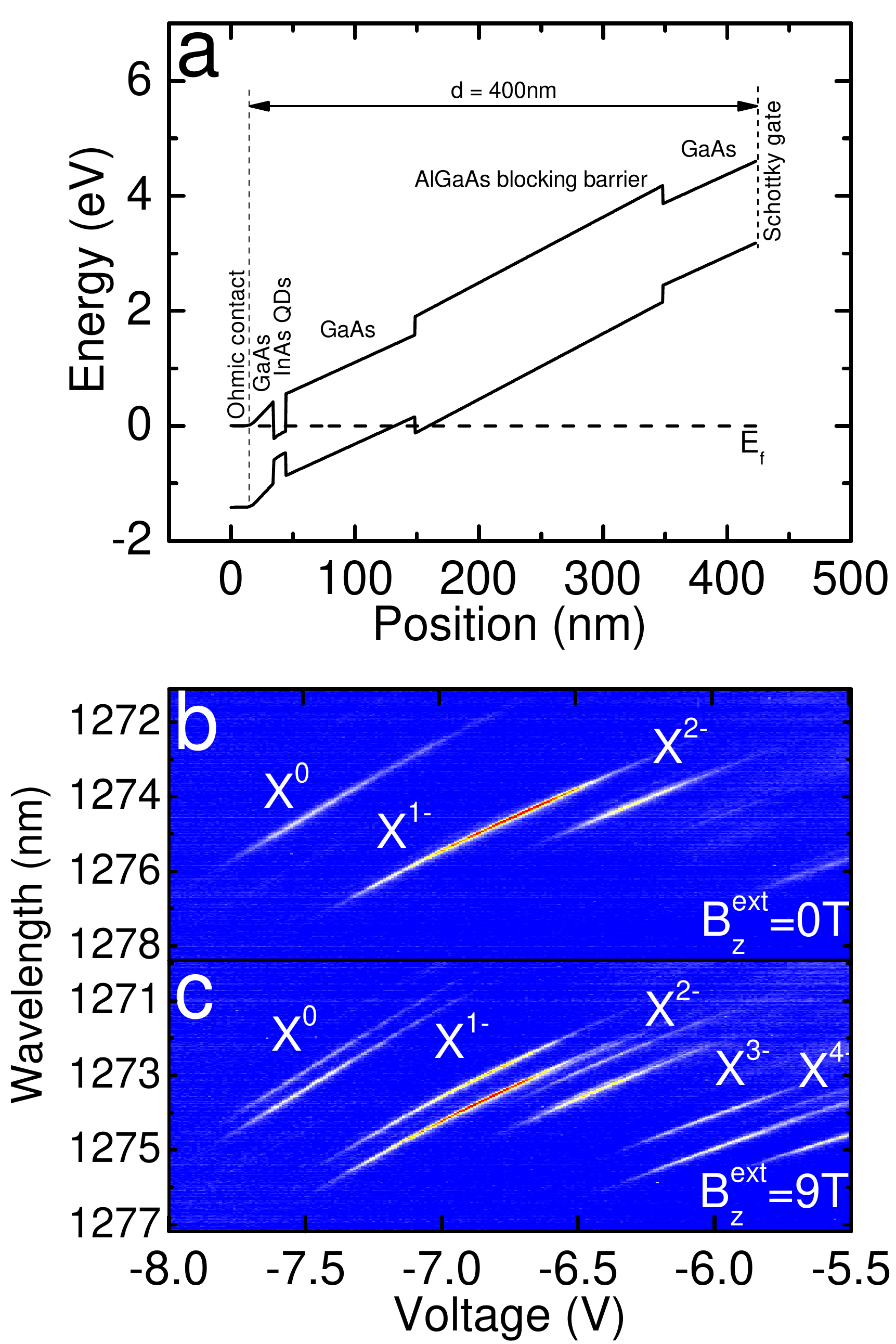}
   \caption{(a) Schematic of the energy diagram of the charge tunable structure under study, including a single layer of telecom wavelength quantum dots (QDs) grown in a quantum well. E$_f$ indicates the Fermi energy level. (b, c) Photoluminescence spectra collected as a function of the applied gate voltage V under non-resonant excitation ($\lambda$ = 830 nm) at T = 4 K. The external magnetic field B applied along the quantum dot growth axis $z$ is 0 T in (b) and 9 T in (c). The emission lines corresponding to the different charge states of a single quantum are labeled accordingly (X$^{0}$ = neutral exciton, X$^{1-}$ = negatively-charged exciton, etc.). }

   \label{Fig1}
\end{figure}

\section{Electric field dependence of single quantum dot confined states}

Examples of photoluminescence spectra acquired as a function of applied voltage are shown in Fig. 1b. Distinct emission lines are visible and can be attributed to single exciton ($X^{0}$) and negatively-charged exciton ($X^{1-}$ and $X^{2-}$) recombinations. Discrete jumps in the emission wavelength, visible when varying the applied voltage, are the signature of the Coulomb blockade effect occuring when an extra electron is added to the quantum dot bound states \cite{charge_tun}. We observe a partial co-existence of excitonic lines of different charged states of the same quantum dot around the transition voltages due to the comparable rates of electron tunneling into the quantum dot from the Fermi sea and exciton recombination (see Fig. 1b) \cite{BDG}.
By applying a perturbative Coulomb blockade model \cite{C_b} to single quantum dots, we can extract important physical parameters including the electron-electron (electron-hole) interaction energy, the electron (hole) confinement energy, and the effective lengths of the electron (hole) wavefunctions. These results are shown in Fig. 2. We find that the Coulomb interaction energies are much smaller than the confinement energies and they can, therefore, be treated as perturbations. Thus, we consider the trapped carriers to be in the so-called strong confinement regime.
Compared to typical 950 nm quantum dots \cite{QD_E, PRB77}, we derive wavefunction effective lengths and confinement energies about a factor two larger and carrier-carrier interaction energies about a factor two smaller. This is compatible with a larger quantum dot physical size. It is worth noting that engineering of larger wavefunction lengths could lead to larger oscillator strengths for the excitonic transitions, relevant for quantum electrodynamics experiments with single quantum dots.
The larger confinement energies have important consequences in the tunability of the transition energies of the quantum dot confined states. While the transition energy of quantum dots emitting below 1 $\mu$m can be tuned by about 1 meV in typical field-effect transistor devices \cite{QD_E}, the telecom wavelength quantum dots under study show a tunability up to 7 meV. This can be explained by the larger electron and hole confinement energies which enable larger electric fields to be applied before the charges tunnel out of the confining potential. This larger wavelength tunability is important for potential applications such as the mutual tuning of the emission lines with respect to optical cavities for quantum optics experiments, or cancellation of the fine structure splitting to create sources of entangled photon pairs by applying an external electric field \cite{FSS_E}. Measurements of the fine structure splitting of quantum dots, from the same growth and within the same structure as the ones presented in this work, were reported in Ref.\cite{PRB}. Further, such large confinement energies yield carriers more decoupled from the electron reservoirs and have potential for reduced impact of phonon-induced dephasing.
When we apply an external bias V, exciton transition energies (E$_{PL}$) experience a Stark shift, following the relation  E$_{PL}$ = E$_0$ - $p$F + $\beta$F$^2$, where the electric field F = -(V$_g$ - V$_0$)/d is a function of the Schottky barrier height V$_0$ and the distance $d$ between the back gate and sample surface, $p$ is the permanent dipole moment and $\beta$ the polarizability. Given the structure of the sample under study (see Fig. 1a), we use V$_g$ = 0.62 V, and $d$ = 400 nm and fit the exciton lines with quadratic functions, as shown in Fig. 3a, to extract the permanent dipole moment $p$ and the polarizability $\beta$.
We observe permanent dipole moments with $p/e$ values (where $e$ is the electron charge) ranging from -0.5 to -3.0 nm, values similar to those reported for quantum dots emitting around 950 nm \cite{QD_E}, indicating that the electron and hole wavefunctions are centered within the quantum dot. The observed polarizabilities (ranging between -0.5 to -1.2 $\mu$eV/(kV/cm)$^2$) are slightly smaller than the values reported for shorter wavelength quantum dots \cite{QD_E}, which can be again explained by the stronger confinement of the carriers in larger telecom wavelength quantum dots. The negative sign of the polarizability implies that the hole is confined near the base of the dot, while the electron wavefunction, given its lighter effective mass, is delocalized over the quantum dot. The polarizability and the permanent dipole moment, as expected \cite{QD_E}, are linearly related, as shown in Fig. 3b.

\begin{figure}[tb!]
   \centering
  \includegraphics[height=19cm, width=1.0\linewidth]{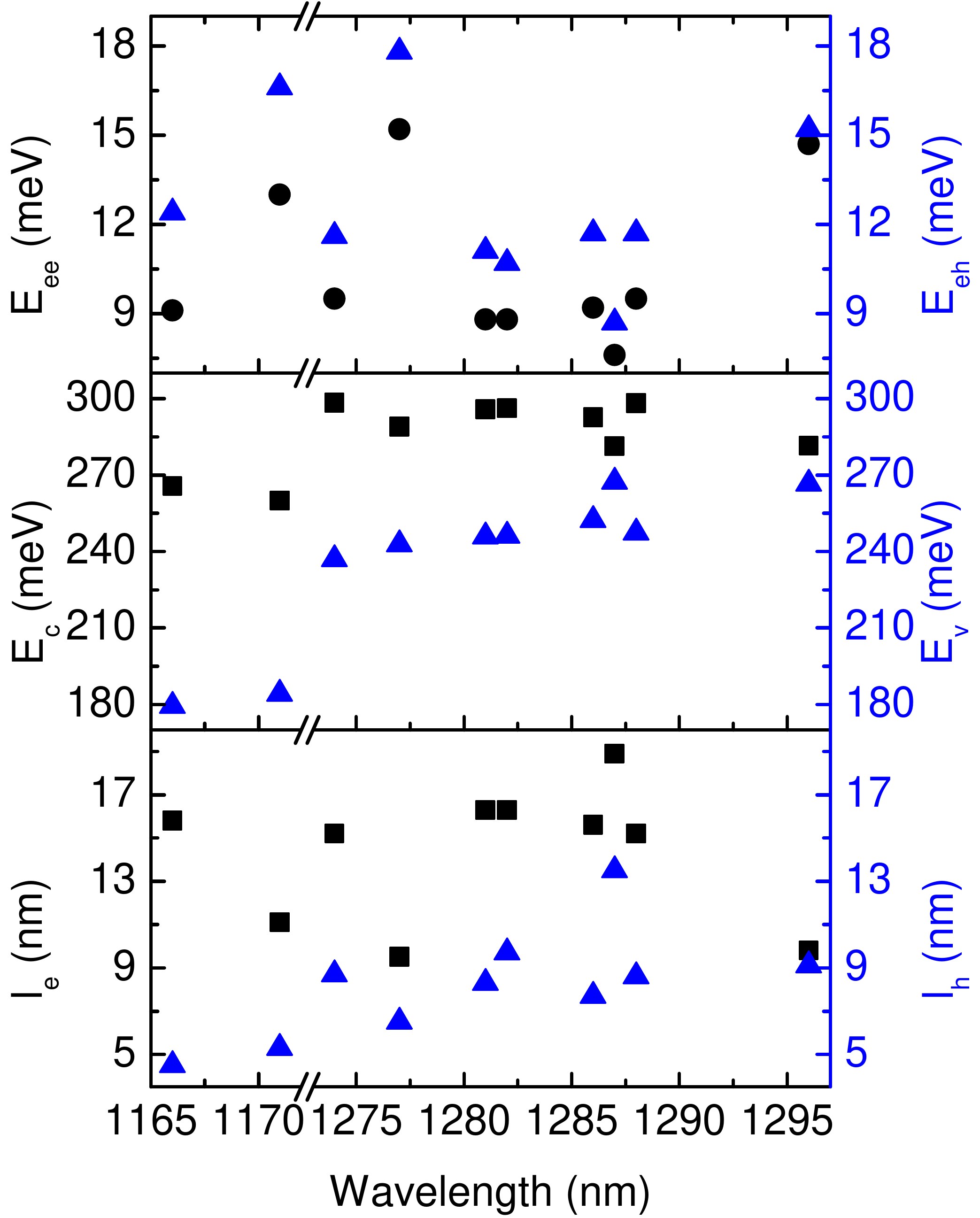}
   \caption{The Coulomb blockade model (see main text) is applied to extract the electron-electron and electron-hole interaction energies ($E_{ee}^{ss}$ and $E_{eh}^{ss}$, respectively), as well as the electron and hole confinement energies ($E_{C}$ and $E_{V}$, respectively) and the electron and hole wavefunction extension ($l_{e}$ and $l_{h}$, respectively). The wavelength on the $x$-axis corresponds to the emission wavelength for the $X^{0}$ line in the middle of the emission tuning range. }

\end{figure}

\begin{figure}[tb!]
   \centering
  \includegraphics[height=9cm, width=0.7\linewidth]{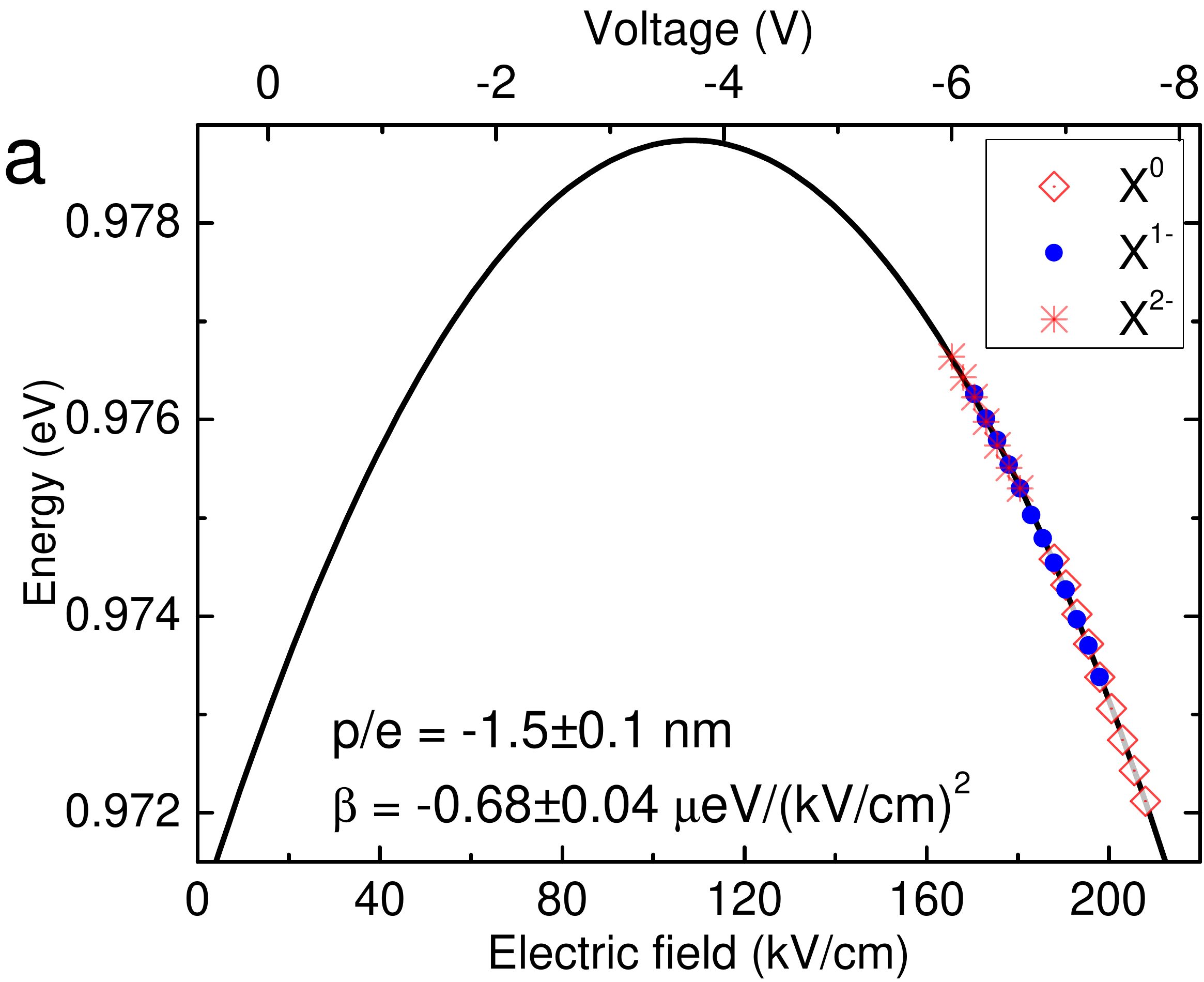}
   \includegraphics[height=9cm, width=0.7\linewidth]{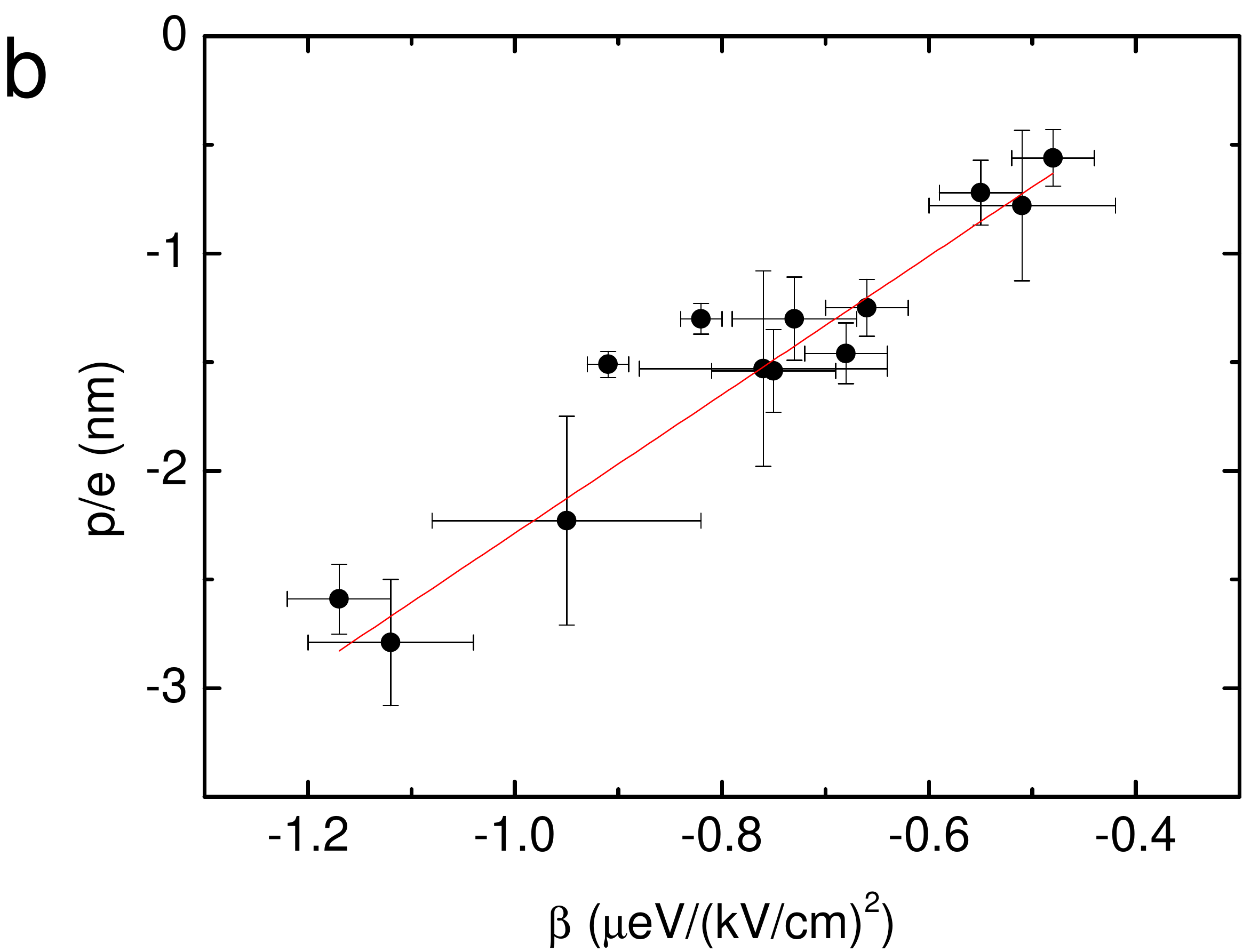}
   \caption{(a) The energies of different charged states from a single quantum dot as a function of the applied electric field. Here the multi-particle Coulomb interaction energies have been subtracted. The solid black line is a parabolic fit to the data. (b) Permanent dipole moments $p$, measured from several single quantum dots, plotted as a function of polarizability $\beta$. The solid red line is a linear fit with slope 3.2 $\pm$ 0.2 nm/(meV/(kV/cm)$^2$) and the error bars are obtained from the errors in the fits.}

   \label{Fig2}
\end{figure}

\section{Investigation of the quantum dot confined states in the presence of an external magnetic field}

To further investigate the properties of the bound-state wavefunctions of the telecom-wavelength quantum dots under study, we apply an external magnetic field \textbf{B} in the Faraday (\textbf{B} parallel to the quantum dot growth direction) and in the Voigt (\textbf{B} orthogonal to the quantum dot growth direction) configurations. Examples of photoluminescence spectra collected at a field of 9 T in the Faraday geometry, when varying the electric field applied to the charge tunable structure are shown in Fig. 1c. A clear splitting of each excitonic transition is visible; this  will be analysed and discussed in more detail in the following sections.

\subsection{Diamagnetic coefficients of excitons in Faraday and Voigt geometries}

We first consider the Faraday geometry, where we apply magnetic fields up to 9 T parallel to the quantum dot growth axis and collect photoluminescence spectra as a function of the voltage applied to the charge tunable structure. Examples of the spectra collected for the negatively- charged exciton are shown in Fig. 4a: the energy of the emission lines  experiences the so-called diamagnetic shift in the presence of an external magnetic field, following the expression E=$\alpha$B$^2$, where $\alpha$ is the diamagnetic coefficient. The values of $\alpha$ obtained from individual quantum dots are shown in Table I. The diamagnetic coefficient is related to the exciton binding and confinement energies, and therefore to the microscopic properties of each specific quantum dot. For the telecom-wavelength quantum dots under study, we generally observe a modest (about 10$\%$) increase of $\alpha$ between the neutral exciton and the charged exciton (see Table I). In the weak confinement regime, due to electron-electron interaction the addition of a second electron to a neutral exciton can reduce $\alpha$ by up to a factor 2 \cite{QD_B}. The fact that we do not observe a reduction, but rather a modest increase, in  $\alpha$  further validates that the carriers are in the strong confinement regime in the quantum dots under study, as reported also in Ref. \cite{Cade}. We observe significant differences between the electron and hole wavefunction extents, as expected due to the difference in confinement potentials and effective masses.
 
By applying the external magnetic field orthogonal to the quantum dot growth axis (Voigt configuration), the rotational symmetry of the wavefunctions is broken. This results in the mixing of the originally bright and dark neutral exciton states, with the latter becoming visible in the photoluminescence spectra \cite{Voigt, Voigt2}. If we consider quantum dots in the 950 nm emission range, the difference between the diamagnetic coefficients measured in Faraday and Voigt configuration reaches values up to about a factor 3 \cite{magnetoPL}. Interestingly, for the telecom wavelength quantum dots under study, the diamagnetic coefficient is one order of magnitude smaller in the Voigt configuration compared to the results obtained in the Faraday configuration. As the diamagnetic coefficient is a measure of the effect of confinement, this striking result confirms the unique morphology of the DWELL quantum dots compared to typical self-assembled quantum dots emitting near 950 nm. As $\alpha$ is an order of magnitude larger for applied fields in-plane versus out-of-plane, we conclude that the confinement in the growth direction is less for DWELL quantum dots as expected.

\begin{figure}[]
  \includegraphics[height=5.5cm, width=0.5\linewidth]{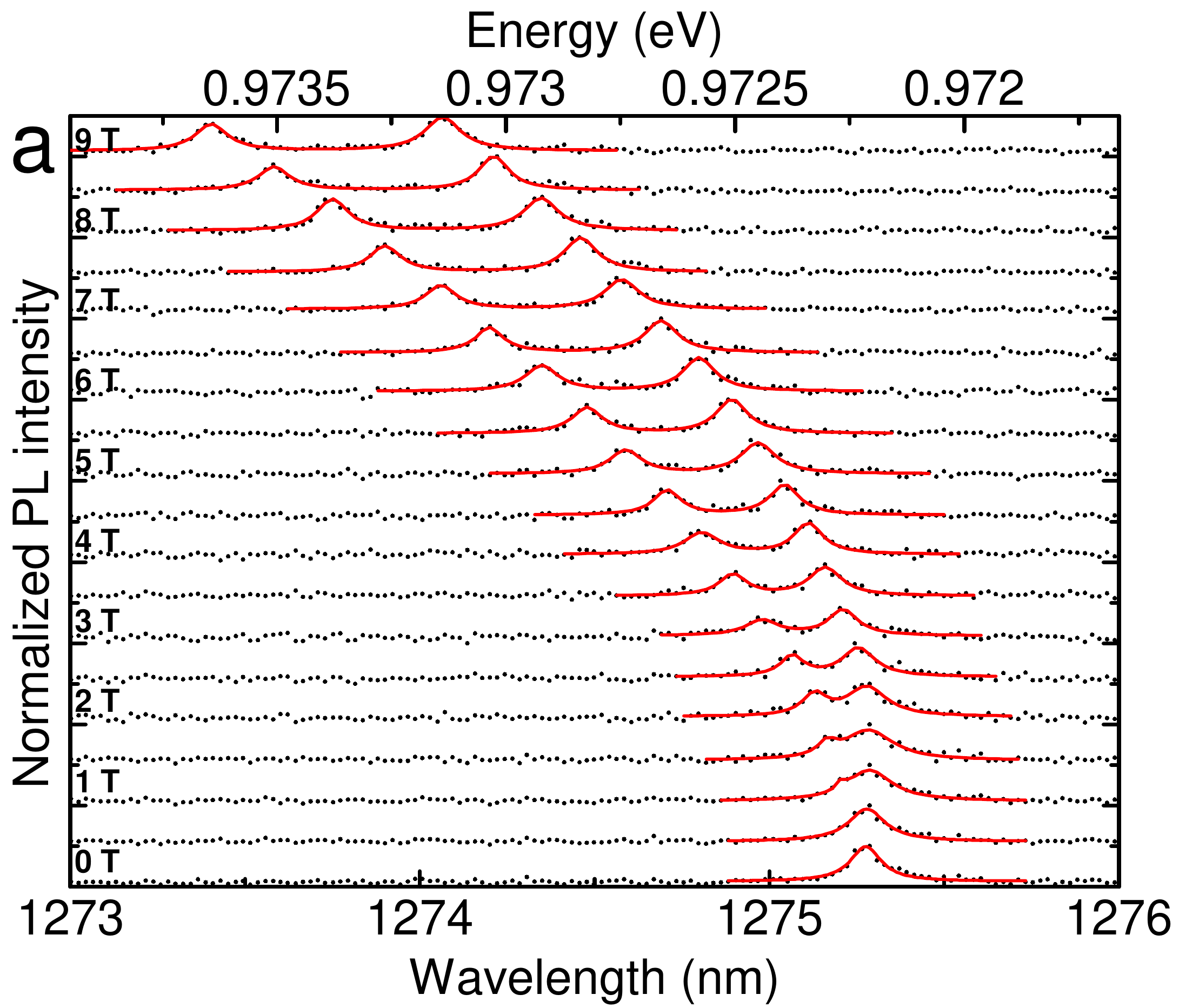}
  \includegraphics[height=5.5cm, width=0.5\linewidth]{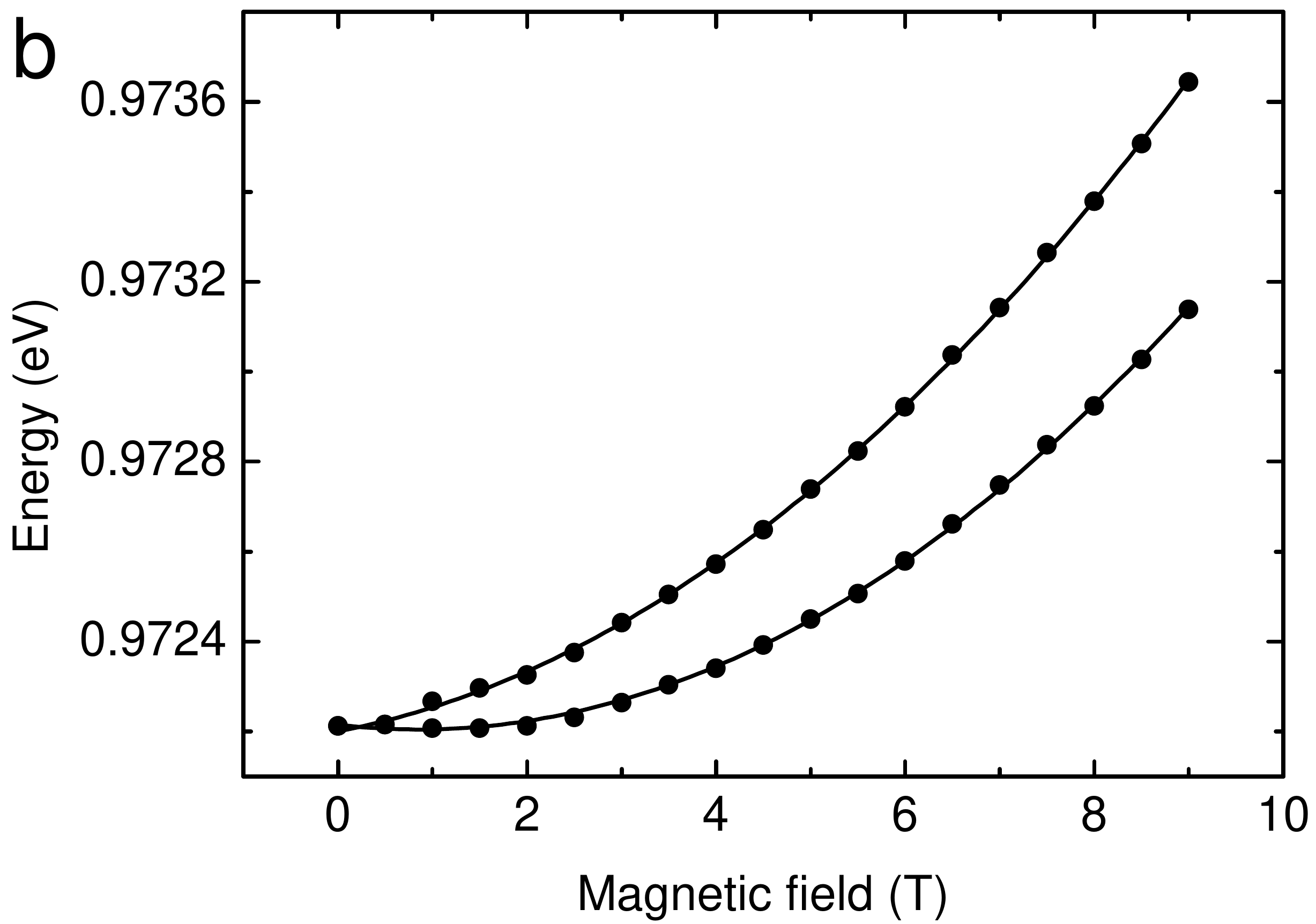}
  \includegraphics[height=5.5cm, width=0.5\linewidth]{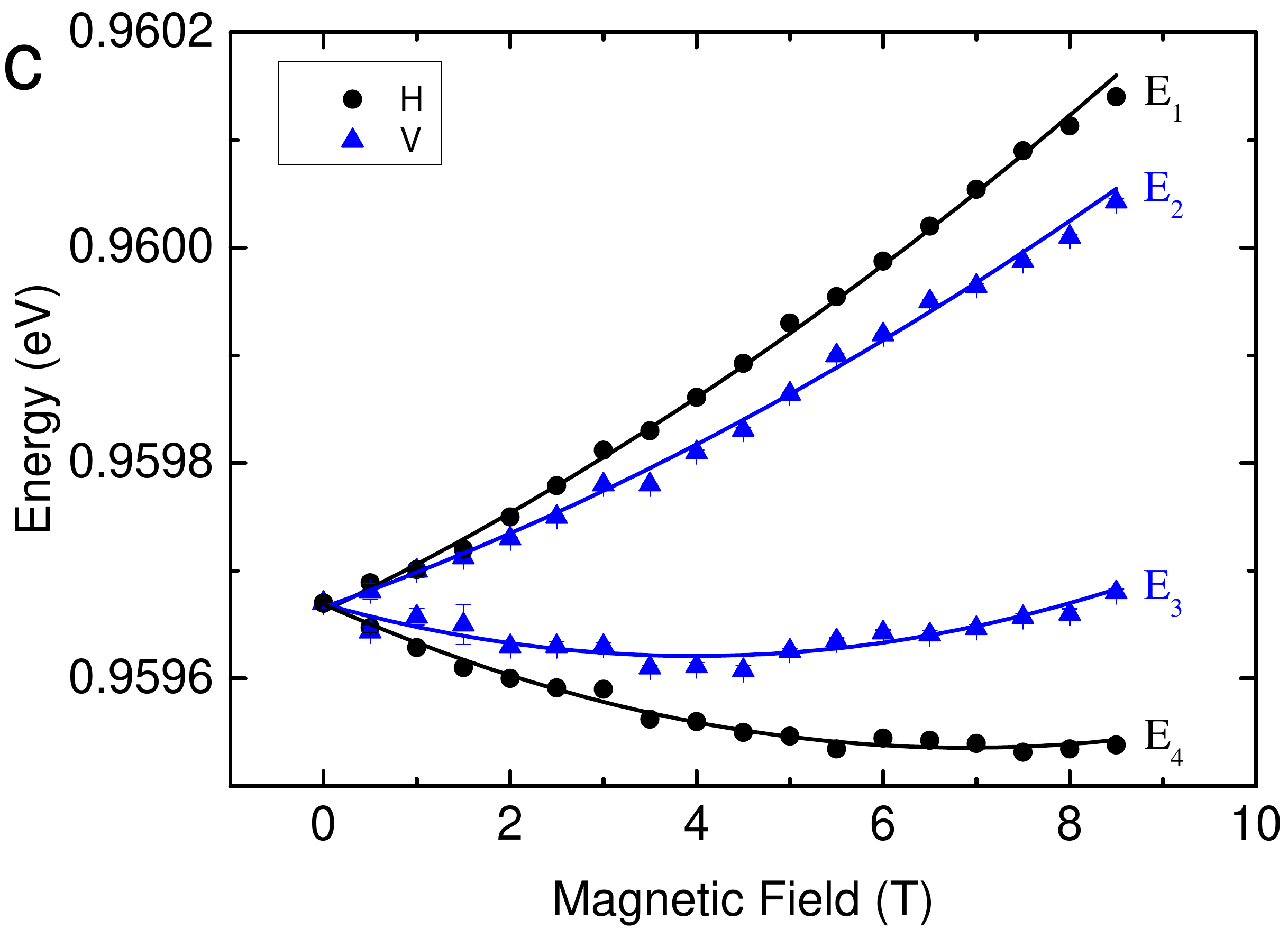}
   \caption{(a) Faraday configuration: Normalised photoluminescence (PL) spectra (shifted for clarity) of a negatively-charged exciton state from a single quantum dot, collected under non-resonant excitation at a temperature of 4K, for applied magnetic fields ranging from 0 T to 9 T (with 0.5 T increments). The solid lines are Lorentzian fits to the data. (b): Energy position of the peaks, as found from the Lorentzian fits of panel (a), plotted as a function of the applied magnetic field. The solid lines are quadratic fits. (c), Voigt configuration: Energy position of the negatively-charged exciton peaks, as found from the Lorentzian fits of the spectra collected for two orthogonal polarisations (horizontal, H and vertical, V), plotted as a function of the applied magnetic field. The error bars (often within the symbol size) in panels (b) and (c) are obtained from the Lorentzian fits of the photoluminescence peaks.}

   \label{Fig3}
\end{figure}

\subsection{Zeeman splitting and g-factors}

Examples of the spectra collected for the negatively-charged exciton in the Faraday configuration are shown in Fig. 4a: the emission line in the presence of the magnetic field is split by the so-called Zeeman splitting (see Fig. 4b) with a magnitude $\Delta$E given by $\Delta$E=$g\mu_{B}$B, where $\mu_{B}$ is the Bohr magneton and $g$ is the Land\'e factor. The Zeeman splittings that we measure range between about 10 and 40 $\mu$eV/T, considerably smaller than for 950 nm quantum dots (that were reported to be 120 $\pm$ 30 $\mu$eV/T \cite{QD_B}). This is consistent with theoretical calculations showing that an increase in the quantum dot size implies a reduction of the $g$-factor \cite{QD_size_g}.

By applying the external magnetic field orthogonal to the quantum dot growth axis (Voigt configuration), the rotational symmetry of the wavefunctions is broken. As shown in Fig. 4c, the single quantum dot negatively-charged exciton line splits into 4 separate contributions (E$_1$, E$_2$, E$_3$, E$_4$), as a result of the Zeeman splitting of the bright and dark states. The s-shell electron and hole $g$-factors can be determined from the exciton energies: by fitting the energies of each transition, one can determine $g_e \mu_B B$= E${_1}$-E${_3}$=E${_2}$-E${_4}$ and $g_h \mu_B B$= E${_1}$-E${_2}$=E${_3}$-E${_4}$ \cite{Shields}. The values that we extract from these measurements are plotted in Table I. We attribute the variations in the values of the $g$-factors for different quantum dots to be due to the dependence of $g$ on the quantum dot shape \cite{Pryor}.

The magnitudes of the $g$-factors measured for the negatively-charged states are consistently higher than the ones measured for the neutral exciton, in accordance with previous reports \cite{Shields}. The electron $g$-factors are two to four times larger than hole $g$-factors since the electron wavefunctions are less confined than the hole ones and are therefore more sensitive to the external magnetic field. From the measurements shown in Fig. 4c, one can also see that the four transitions are linearly polarised and, as expected, two of the four transitions disappear when polarisation is resolved into horizontal and vertical components.

\section{Conclusions}

In summary, we have fully characterized the exciton and carrier properties of single quantum dots emitting at telecom wavelengths (near 1.3 $\mu$m) under applied electric and magnetic fields.  Via the  Coulomb-blockade model, we extract the electron and hole wavefunction lengths as well as multi-particle Coulomb interaction and confinement energies. The results are consistent with a strong-confinement picture. The confinement energies of these quantum dots are found to be a factor of two larger than 950 nm quantum dots. Due to the larger confinement energies, the excitonic transitions can be tuned over a larger range than 950 nm band quantum dots in comparable devices. Additionally, the deeper confinement holds promise for better decoupled spins from Fermi or phonon reservoirs. With applied external magnetic fields we extract the Zeeman and diamagnetic coefficients as well as electron and hole $g$-factors. Compared to 950 nm quantum dots, the Zeeman splittings are significantly smaller and the diamagnetic coefficient shows a drastic shift when changing from the Faraday to the Voigt configuration due to the unique DWELL morphology. These results give insights into the fundamental properties of telecom wavelngth quantum dots. Further investigations into the impact of the DWELL morphology on the strain in the dot (and its  effect e.g. on the heavy-hole/light-hole mixing in the valence band and the electron and hole spin coupling to nuclear spins) are needed to validate the promising potential of telecom-wavelength quantum dots for future quantum information applications.

\newpage

\begin{table}[htb!]
\centering
\begin{tabular}{ |P{4cm}|P{4cm}|P{4cm}|P{4cm}|  }
 \hline
 \multicolumn{4}{|c|}{\textbf{Faraday configuration}} \\
 \hline

 $\lambda_{X_0}$(nm)  &   Excitonic state  &    $\mid g \mid$  &    $\alpha$($\mu$eV/T$^2$) \\
 \hline
  1274   & X$^0 $    &0.73$\pm$0.02& 13.48$\pm$0.09\\
    & X$^{1-}$    &0.63$\pm$0.01& 14.83$\pm$0.07\\
 \hline
 1281   & X$^0 $    &0.77$\pm$0.01& 17.01$\pm$0.20\\
    & X$^{1-}$    &0.89$\pm$0.01& 18.20$\pm$0.31\\
 \hline
  1282   & X$^0 $    &0.36$\pm$0.02& 14.25$\pm$0.09\\
    & X$^{1-}$    &0.98$\pm$0.01& 14.72$\pm$0.04\\
 \hline
 \end{tabular}
 
 \begin{tabular}{ |P{4cm}|P{4cm}|P{1.92cm}|P{1.92cm}|P{4cm}|  }
 \multicolumn{5}{|c|}{\textbf{Voigt configuration}}\\
 \hline
 $\lambda_{X_0}$(nm)  &    Excitonic state  &    g$_h$  & g$_e$  &   $\alpha$($\mu$eV/T$^2$) \\
 \hline
 1290   & X$^0 $    &-0.17$\pm$0.01&-0.77$\pm$0.01 & 2.15$\pm$0.12\\
    & X$^{1-}$    &-0.49$\pm$0.06&-0.91$\pm$0.07 & 5.30$\pm$0.47\\
 \hline
 1292   & X$^{1-}$    &-0.24$\pm$0.01&-1.04$\pm$0.01 & 2.50$\pm$0.05\\
   
 \hline
 1300   & X$^0 $    &-0.21$\pm$0.02&-0.87$\pm$0.04 & 0.93$\pm$0.30\\
    & X$^{1-}$    &-0.36$\pm$0.02&-0.80$\pm$0.03 & 1.83$\pm$0.26\\
    \hline
 1310   & X$^{1-}$    &-0.26$\pm$0.04&-0.89$\pm$0.04 & 1.82$\pm$0.19\\
 \hline
 1317   & X$^{1-}$    &-0.19$\pm$0.05&-0.72$\pm$0.03 & 0.63$\pm$0.14\\
  
 \hline

\end{tabular}
  \caption{Diamagnetic coefficients $\alpha$, $g$-factors and electron and hole $g$-factors ($g_e$ and $g_h$, respectively) extracted from single quantum dot photoluminescence spectra collected under external magnetic field applied in the Faraday and Voigt configurations. $\lambda_{X_0}$ corresponds to the emission wavelength of the excitonic line in the middle of the emission tuning range.}

\end{table}


\begin{thebibliography}{100}

\bibitem{Reiz_ind_photons} M. Gschrey et al, Nature Communications \textbf{6}, 7662 (2015).

\bibitem{Ding} X. Ding et al., Phys. Rev. Lett. \textbf{116}, 020401 (2016).

\bibitem{entang} R. J. Young et al., New Journal of Physics \textbf{8}, 29 (2006).

\bibitem{entang2} A. Delteil, Z. Sun, W. Gao, E. Togan, S. Faelt, A. Imamoglu, Nature Physics \textbf{12}, 218 (2016).

\bibitem{QDspin_review} R. J. Warburton, Nature Materials \textbf{12}, 483 (2013).

\bibitem{sup_det} R. H. Hadfield, Nature Photonics \textbf{3}, 696 (2009). 

\bibitem{sup_det2} F. Marsili et al., Nature Photonics \textbf{7}, 210 (2013).

\bibitem{Edo} J.-H. Kim et al., http://arxiv.org/abs/1511.05617

\bibitem{Shields_inter} M. Felle et al., Appl. Phys. Lett. 107, 131106 (2015).

\bibitem{Shields_entang} M.B. Ward et al., Nature Communications \textbf{5}, 3316 (2014).

\bibitem{PRB} L. Sapienza et al., Phys. Rev. B \textbf{88}, 155330 (2013).

\bibitem{DWELL} K. Srinivasan, O. Painter, A. Stintz, and S. Krishna, Appl. Phys. Lett. \textbf{91}, 091102 (2007).

\bibitem{DWELL1} K. Nishi et al., Appl. Phys. Lett. \textbf{74}, 1111 (1999).

\bibitem{DWELL2} V. M. Ustinov et al., Appl. Phys. Lett. \textbf{74}, 2815 (1999).

\bibitem{laser} D. L. Huffaker et al., Appl. Phys. Lett. \textbf{73}, 2564 (1998).

\bibitem{QD} B. Alloing et al., Appl. Phys. Lett. \textbf{86}, 101908 (2005).

\bibitem{1550_B} V. V. Belykh et al., Phys. Rev. B \textbf{92}, 165307 (2015).

\bibitem{Kleemans} N.A.J.M. Kleemans et al., Phys. Rev B \textbf{79}, 045311 (2009).

\bibitem{Cade} N. I. Cade et. al. Phys. Rev. B \textbf{73}, 115322 (2006).

\bibitem{charge_tun} R.W. Warburton et al., Nature \textbf{405}, 926 (2000).

\bibitem{SIL} K. A. Serrels et al., Journal of Nanophotonics \textbf{2}, 021 854 (2008).

\bibitem{BDG} B.D. Gerardot et al., Appl. Phys. Lett., \textbf{99}, 243112 (2011).

\bibitem{C_b} R. J. Warburton et al., Phys. Rev. B \textbf{58}, 16221 (1998).

\bibitem{QD_E} R. J. Warburton et al., Phys. Rev. B \textbf{65}, 113303 (2002).

\bibitem{PRB77} P. A. Dalgarno et al., Phys. Rev. B \textbf{77}, 245311 (2008).

\bibitem{FSS_E} A. J. Bennett et al., Nature Physics \textbf{6}, 947 (2010).

\bibitem{QD_B} C. Schulhauser et al., Phys. Rev. B \textbf{66}, 193303 (2002).

\bibitem{Voigt2} M. Bayer et al., Phys. Rev. B \textbf{65}, 195315 (2002).

\bibitem{Voigt} M. Bayer, O. Stern, A. Kuther, and A. Forchel, Phys. Rev. B \textbf{61}, 7273 (2000).

\bibitem{magnetoPL} B. van Hattem et al., Phys. Rev. B \textbf{87}, 205308 (2013).

\bibitem{QD_size_g} R. Zielke et al., Phys. Rev. B \textbf{89}, 115438 (2014).





\bibitem{Shields}  A. J. Bennett et al., Nat. Commun. \textbf{4}, 1522 (2013).

\bibitem{Pryor} C. E. Pryor et al., Phys. Rev. Lett. \textbf{96}, 026804 (2006).

\bibitem{nanowires_B} B. J. Witek et al., Phys. Rev. B \textbf{84}, 195305 (2011).



\end{thebibliography}
\end{document}